\documentclass[fleqn,usenatbib]{mnras}

\usepackage{newtxtext,newtxmath}

\usepackage[T1]{fontenc}

\DeclareRobustCommand{\VAN}[3]{#2}
\let\VANthebibliography\thebibliography
\def\thebibliography{\DeclareRobustCommand{\VAN}[3]{##3}\VANthebibliography}

\usepackage{graphicx}	
\usepackage{amsmath}	
\usepackage{siunitx}

\newcommand{\lmc}{LMC X--4}
\newcommand{\astro}{\emph{AstroSat}}
\newcommand{\RXTE}{\emph{RXTE}}
\newcommand{\nus}{\emph{NuSTAR}}
\newcommand{\xmm}{\emph{XMM}-Newton}

\title[Orbital ephemerides of \lmc]{A comprehensive study of orbital evolution of \lmc: Existence of a second derivative of the orbital period}

\author[C. Jain, R. Sharma \& B. Paul]{
Chetana Jain,$^{1}$\thanks{E-mail: chetanajain11@gmail.com (CJ)}
Rahul Sharma,$^{2}$\thanks{E-mail: rahul1607kumar@gmail.com (RS)}
and Biswajit Paul$^{2}$
\\
$^{1}$Hansraj College, University of Delhi, Delhi 110007, India\\
$^{2}$Astronomy \& Astrophysics, Raman Research Institute, C.V. Raman Avenue, Bangalore 560080 Karnataka, India
}

\date{Accepted XXX. Received YYY; in original form ZZZ}

\pubyear{2024}

\begin{document}
\label{firstpage}
\pagerange{\pageref{firstpage}--\pageref{lastpage}}
\maketitle

\begin{abstract}

We report here results from pulse arrival time delay analysis of the eclipsing high mass X-ray binary pulsar \lmc\ using observations made with the Rossi X-ray Timing Explorer, \xmm, \nus\ and \astro. Combining the orbital parameters determined from these observations with the historical measurements dating back to 1998, we have extended the $T_{\pi/2}$ epoch history of \lmc\ by about 4600 binary orbits spanning about 18 years. We also report mid-eclipse time measurements ($T_{ecl}$) using data obtained from wide-field X-ray monitors of \textit{MAXI}-GSC and \textit{Swift}-BAT. Combining the new $T_{\pi/2}$ and $T_{ecl}$ estimates with all the previously reported values, we have significantly improved the orbital evolution measurement, which indicates that the orbital period is evolving at a time scale ($P_{\rm orb}/\dot{P}_{\rm orb}$ ) of about 0.8 Myr. For the first time in an accreting X-ray pulsar system, we confirm the existence of a second derivative of the orbital period, having an evolution time scale ($\dot{P}_{orb}/\ddot{P}_{orb}$) of about 55 yr. Detection of a second derivative of the orbital period in \lmc\ makes its orbital evolution timescale more uncertain, which may also be true for other HMXBs. Independent solutions for the orbital evolution measurement using the mid-eclipse data and the pulse timing data are consistent with each other, and help us put an upper limit of 0.009 on the eccentricity of the binary system.
\end{abstract}




\section{Introduction}

It has been long realized that most of the bright galactic X-ray sources occur in binary systems \citep{Verbunt93, Sana12}, and binary evolution plays a key role in understanding their stellar evolution \citep{Heuvel94}. Several key astrophysical phenomena, such as the formation of double compact binary (double black holes, black hole-neutron star and double neutron star systems), followed by the merger of the two stellar components, production of short gamma-ray bursts and eventually a possible gravitational wave detection requires a comprehensive understanding of the interaction between the binary components \citep{Belczynski02}. Measurements of the orbital period decay are often used to place limits on the mass transfer and/or rate of mass loss from the system \citep{Deeter81}. Orbital decay of the famous Hulse-Taylor binary pulsar provided an unprecedented insight into the loss of energy due to gravitational waves, consistent with the general theory of relativity \citep{Taylor82}. 

Being progenitors of double compact objects, the orbital period of X-ray binaries and their evolution has been extensively studied. Important ingredients of such studies include the effect of mass exchange between binary components and mass loss from the binary system on the orbital parameters. The orbital evolution mechanisms largely include mass transfer (conservative as well as non-conservative) from the companion star to the compact object \citep{Heuvel94}, tidal dissipation in close binaries \citep{Lecar76, Zahn77}, loss of orbital angular momentum due to stellar winds \citep{Brookshaw93}, gravitational wave radiation \citep{Verbunt93} and magnetic activity associated with the companion star \citep{Wolff09, Jain11, Jain22}.

The current work is an accurate and most up-to-date study of the orbital evolution of \lmc\ which is an eclipsing high mass X-ray binary (HMXB) system located about 50 kpc away in the Large Magellanic Cloud (LMC) \citep{Giacconi72}. It was discovered in 1972 by the UHURU observatory. This system comprises of a 1.25 M$_{\odot}$ neutron star in an almost circular orbit around a 14th magnitude OB star \citep{Pakull76, Pesch76, Kelley83, Meer07}. The timing variability of \lmc\ includes 13.5 s coherent pulsations \citep{Kelley83}, $\sim$26 mHz quasi-periodic oscillations \citep{Rikame2022, Sharma23}, about an hour long flaring episodes \citep{Epstein77, Skinner80, Levine91, Beri2017}, about 5 hr long X-ray eclipse \citep{White78}, an orbital period of $\sim$1.4 d \citep{Li78, White78} decaying at a rate of about 10$^{-6}$ yr$^{-1}$  \citep{Falanga15} and 30.5 d intensity variation due to a precessing tilted accretion disk \citep{Lang81, Paul2002b, Molkov2015}.

The orbital evolution in eclipsing X-ray binary pulsars is measured using two well established techniques, viz., timing of the X-ray eclipses and the measurement of the orbital epoch using pulse arrival time delays across the orbital phases. The eclipse timing technique is based on the hypothesis that since the binary components revolve around their center of mass in Keplerian orbits, they are expected to eclipse each other after regular intervals of time. But if the orbit of the binary is perturbed, then the occurrence of eclipses is delayed (increasing orbit) or is advanced (decreasing orbit) in time. Therefore, in this technique, the mid-eclipse times ($T_{ecl}$) are measured over a sufficiently long time base and time connecting them gives estimates on the orbital period evolution of the system. This method has been used to determine the orbital evolution in several eclipsing X-ray binaries, such as, 4U 1822--37 \citep{Jain10b}, AX J1745.6--2901 \citep{Ponti2017}, EXO 0748--676 \citep{Wolff09}, XTE J1710-281 \citep{Jain11, Jain22},  MXB 1658--298 \citep{Jain17}, 4U 1700--37 \citep{Rubin96, Falanga15, Islam2016} etc.

The pulse arrival time technique \citep{Staubert09} is based on correcting the arrival time of the pulses for the binary motion of the compact object. In this method, the pulse profiles are produced using the already known pulse period, P$_{\rm spin}$ \citep[using $\chi^2$ maximisation and epoch folding; ][]{Leahy83}. Assuming $t_0$ is the reference time of the first pulse, and considering non-zero first derivative of the pulse period, ignoring its higher derivatives, the expected arrival time of the $n^{\rm th}$ pulse ($t_n$) as a function of the pulse number (\textit{n}), is given by Equation~\ref{eqn:spin2},
\begin{equation}
t_n = t_0 + P_{\rm spin} n + \frac{1}{2} \dot{P}_{\rm spin} P_{\rm spin} n^2 + a_x \sin{i} \cos{2\pi\left(\frac{t-T_{\pi/2}}{P_{orb}}\right)}
\label{eqn:spin2}
\end{equation}
Assuming an almost circular orbit (eccentricity, $e<<1$), the fourth term in this equation corresponds to the pulse arrival time delay due to the orbital motion. Here, $a_x \sin{i}$ is the projected radius of the orbit and $T_{\pi/2}$ is the mean orbital longitude of the neutron star and corresponds to the maximum delay in the pulse arrival time. An important limitation of this method is the fact that this technique requires sufficiently long observations, covering at least a significant part of the binary orbit. This method has been used in X-ray binaries, such as SAX J1808.4--3658 \citep{Jain07, Burderi09}, Her X-1 \citep{Deeter91, Staubert09}, 4U 1538-52 \citep{Baykal06, Mukherjee06}, Cen X-3 \citep{Kelley83b, Raichur10}, SAX J1748.9-2021 \citep{Sanna2016, Sharma2020} etc.

The first measurements of the pulse arrival time delay in \lmc\ were reported by \citet{Kelley83} using the SAS-3 observations. Later, using the pulse timing analysis, \citet{Dennerl91} and \citet{Levine91} established upper limits of $\sim10^{-6}$ yr$^{-1}$ on the orbital decay. \citet{Harb96} and \citet{Woo96} discussed the orbital period decay in \lmc\ in the context of conservative mass transfer and tidal evolution, superposed by mass loss from the binary system in the form of stellar winds. Using a long \RXTE\ observation of 1998 October, \citet{Levine00} gave the first definite estimate of the decay rate which was refined by \citet{Naik04}, \citet{Falanga15} and \citet{Molkov2015} using data fetched from several X-ray missions.

In this paper, we present an update on the orbital parameters of \lmc\ using both the methods described above. Historically, the orbital evolution measurements of \lmc\ have been done using one of these methods or by combining them; but using only a partial set of the available data. For example, \citet{Falanga15} used only the eclipse data, \citet{Levine00} and \citet{Naik04} used only pulse arrival time data and \citet{Molkov2015} used a part of available data from both techniques.

The paper is organized in the following way. Section §2 gives a description of the observation and the data reduction procedure. In Section §3, the results from the timing analysis of \lmc\ are presented. Our findings are discussed in Section §4.

\section{OBSERVATIONS}

The observation details of the narrow field instruments from which data has been used for the current work are given in Table~\ref{table:obs}. For all the data analysed in this work, the source position (R.A. (J2000) $=05^h 32^m 49.555^s$ and dec. (J2000) $=-66^{\circ} 22' 13.202''$) was adopted from \citep{Gaia2020} to convert the photon arrival times to the solar system barycenter. As described in the previous section, short observations of \lmc\ have not been used in the current work for pulse timing analysis. All the light curves were extracted with a bin time of 0.1 s.

\lmc\ was observed with Rossi X-Ray Timing Explorer \citep[RXTE:][]{Bradt93} several times between 1996 to 1999, amongst which \citet{Levine00} analysed the longest ($\sim16$ day long) observation of 1998 October to obtain a definite measurement of the orbital decay. For the current work, we have used data from the observation made with the Proportional Counter Array (PCA: \citet{Jahoda96} in 1996 August (observation ID: 10135-01) and 1999 December (Observation ID: 40064-01). The total span for both of these observations was more than the orbital period of \lmc\ and for each observation, a useful exposure of $\sim$60 ks was obtained. We have analysed data collected in the Good Xenon mode having a time resolution of 1 $\mu$s. The combined 2--20 keV light curve was extracted using the \textsc{seextract} tool of XRONOS sub-package of \texttt{FTOOLS} \citep{Blackburn99}. The background light curve was extracted from the Standard--2 mode data by using \textsc{pcabackest} with a bright source background model (pca\_bkgd\_cmbrightvle\_eMv20051128). Since a different number of PCUs were operational during both the \RXTE~ observations, therefore, we used the \textsc{correctlc} tool to calculate the equivalent count rate for the simultaneous operation of all five PCUs. The time series was barycenter corrected by using \textsc{faxbary}.

\lmc\ was observed six times with \xmm\ \citep{Jansen01}, out of which, \citet{Molkov2015} have reported $T_{ecl}$ measurements from the first two observations. For the current work, we have analysed the observation of 2003 September which had the longest exposure time ($\sim55$ ks) covering about 45\% of the binary orbit. \xmm\ carries three focal plane European Photon Imaging Cameras (EPIC) for three X-ray telescopes, EPIC-MOS1, -MOS2 and -pn \citep{Struder01, Turner01}. The raw data files were processed using version 20.0.0 of the XMM Science Analysis System (SAS). For the present analysis, we have used 0.5--10 keV data taken with the EPIC-pn detector. The source events were extracted from a circular region of radius \SI{40}{\arcsecond} centred on the source position. The background events were extracted from a similar region centred away from the source location. The event arrival times in the background subtracted light curve were corrected to the solar system barycenter using the SAS tool \textsc{barycen}.

The Nuclear Spectroscopic Telescope ARray \citep[\nus: ][]{Harrison13} is a focusing high-energy X-ray telescope which operates in the 3--79 keV energy band. It comprises of two identical focal plane modules (focal plane modules A (FPMA) and B (FPMB)). For the current work, \nus\ observation of July 2012 was used with a duration of 62 ks covering $\sim$50\% of the binary orbit. We have used the standard \nus\ analysis software \texttt{NuSTARDAS} and the latest calibration files (version 20230124) for data reduction and analysis. The clean event files were generated using the \textsc{nupipeline}. Source events were extracted from a circular region of radius \SI{100}{\arcsecond} centred at the source position. The background events were extracted from a similar region away from the source location. Barycenter correction was done using \textsc{barycorr}. The light curves from both detectors were added for further analysis.

\astro\ is India's first multi-wavelength (from optical to hard X-rays) astronomical mission. It was launched by Indian Space Research Organization in September 2015 \citep{Agrawal06}. Large Area X-ray Proportional Counters \citep[LAXPC;][]{Agrawal17} is one of the primary instruments onboard the \astro. It has a high time resolution of 10 $\mu$s and covers a broad X-ray spectral band in 3--80 keV. It consists of three co-aligned proportional counters (LAXPC10, LAXPC20 and LAXPC30), with a total effective area of 6000 cm$^{-2}$ at 15 keV.  The \astro\ observation of \lmc\ (Observation Id G05\_115T01\_9000000634) made during 2016 August was analyzed for this work. Detailed analysis of this observation has been presented in \citet{Sharma23}. For the current analysis, we used data from LAXPC10 and LAXPC20. LAXPC30 was not used due to gain variability with the instruments \citep{Agrawal17, Antia17}. During this observation, \lmc\ was observed for a duration of $\sim$90 ks, covering about 75\% of the binary orbit. The Event Analysis (EA) mode data from LAXPC10 and LAXPC20 was processed by using the standard LAXPC software\footnote{\url{http://astrosat-ssc.iucaa.in/?q=laxpcData}} (\textsc{LaxpcSoft}: version 3.4.2). The light curves for the source and background from both LAXPC units were extracted from level 1 files by using the tool \textsc{laxpcl1} and added using \textsc{lcmath}. 

\begin{table}
\centering
\caption{Log of X-ray observations of \lmc\ used in this work.}
\label{table:obs}
\resizebox{\columnwidth}{!}{
\begin{tabular}{l cccc}
\hline
Observatory    & Observation & Date & Observation & Exposure$^a$ \\ 
   &   ID  & YY-MM-DD & span (ks) & (ks) \\
\hline
\RXTE & P10135-01 & 1996-08-19 & 132 & 64 \\
\RXTE & P40064-01 & 1999-12-19 & 150 & 60 \\
\xmm & 0142800101 & 2003-09-09 & 113 & 55 \\
\nus & 10002008001  &  2012-07-04 & 62 & 40  \\ 
\astro & 9000000634 & 2016-08-29 & 89 & 31.5 \\
\hline
\multicolumn{5}{l}{$^a$ Net exposure after removing flaring, eclipse and orbital gaps, if any.} \\
\end{tabular}}
\end{table}

\section{TIMING ANALYSIS}

\begin{figure*}
  \centering
  \includegraphics[width=0.9\linewidth]{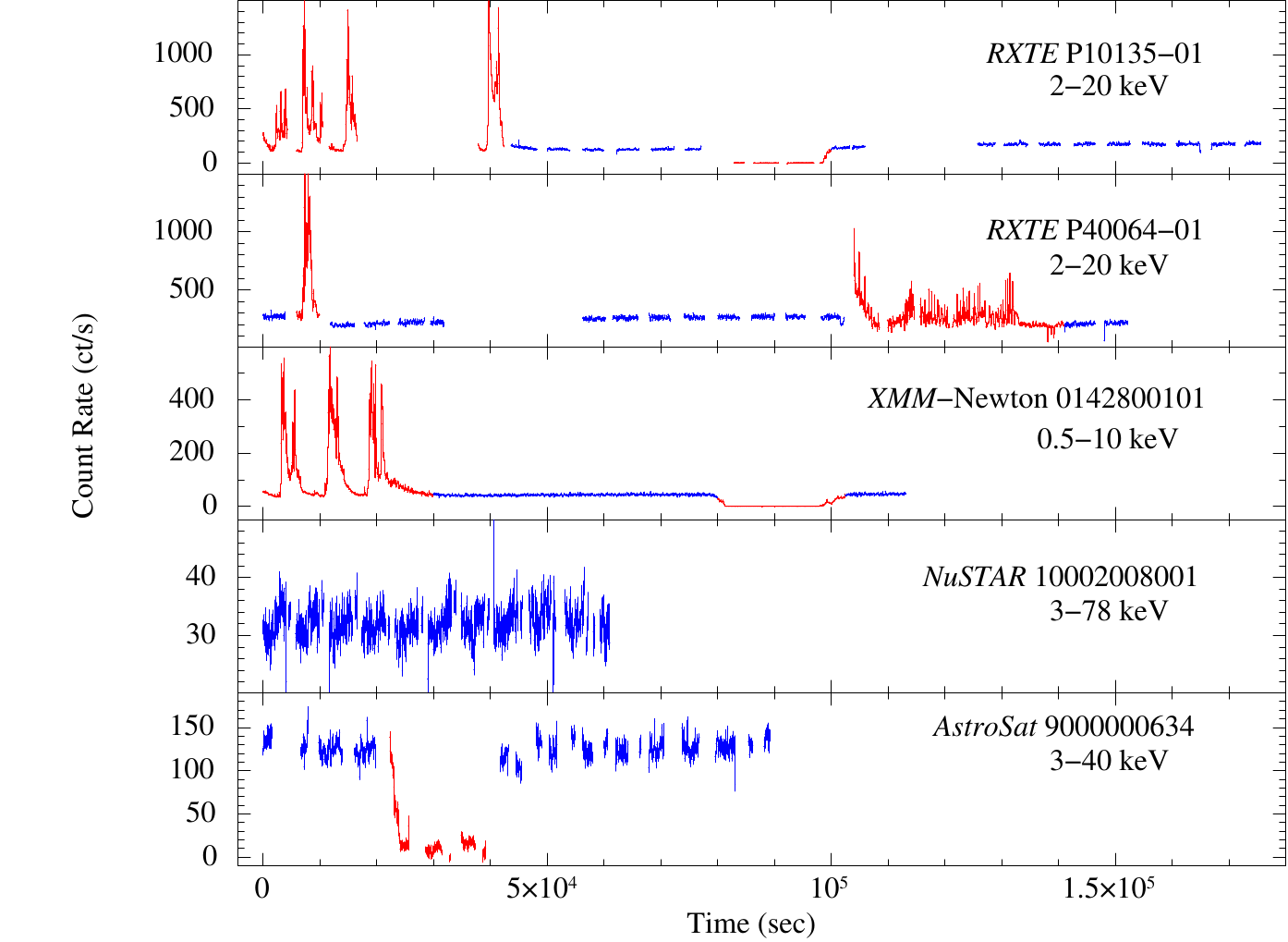}
  \caption{The light curve of \lmc\ obtained from all the observations listed in Table~\ref{table:obs}. The respective X-ray mission, observation ID and energy range} are mentioned in each panel. The blue colored segments were used for the pulse timing measurements. The segments shown in red color were not used. 
  \label{fig:lc}
\end{figure*}

The background subtracted and barycenter corrected light curves of observations listed in Table~\ref{table:obs} were filtered for the flaring and eclipse phases. Figure~\ref{fig:lc} shows all these five light curves to highlight the segments used and those that were filtered off for the pulse timing measurements.

\subsection{$T_{\pi/2}$ measurements}

\lmc\ has an orbital period of $\sim$1.4 d and semi-amplitude of the arrival time delay due to orbital motion ($a_x \sin i$) of $\sim$26.3 s. As a result, the pulse frequency gets modulated by the Doppler effect associated with the orbital motion and the pulses are expected to lose coherence within a few thousand seconds. Assuming a nearly circular orbit, we corrected the photon arrival time for the binary motion. The value of $a_x \sin i$ was fixed to the value taken from \citet{Levine00} and for each observation, $P_{\rm orb}$ was calculated by extrapolating the orbital solution of \citet{Molkov2015}.  

\begin{figure}
  \centering
  \includegraphics[width=0.9\linewidth]{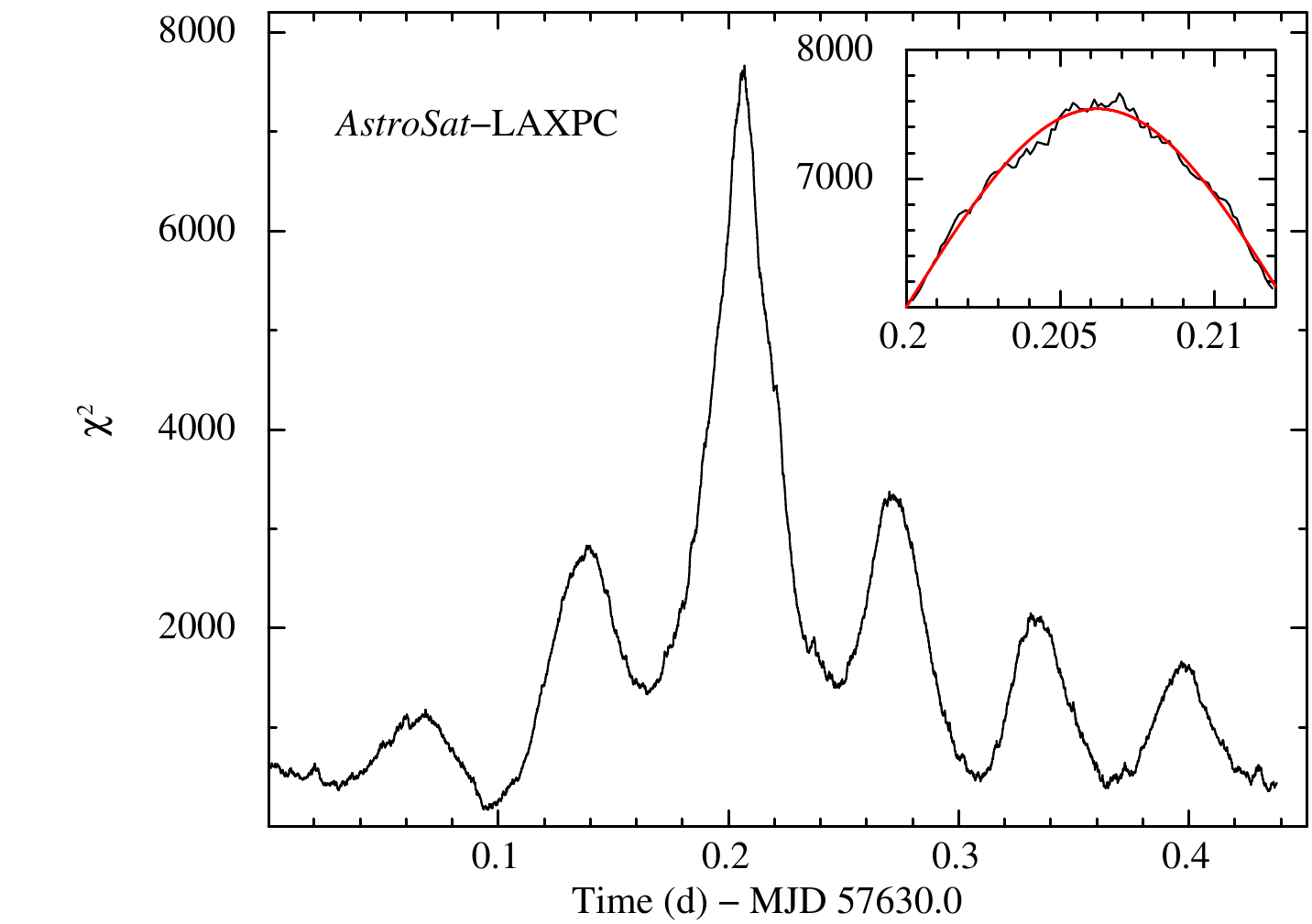}
  \caption{The variation of $\chi^2$ as a function of trial ephemeris for the \astro\ observation of \lmc. \textit{Inset figure:} The solid red line shows the best fit Gaussian profile.}
  \label{fig:t0_Chisq}
\end{figure}

For all the pointed observations of \RXTE, \xmm, \nus\ and \astro\ tabulated in Table~\ref{table:obs}, we searched for the correct orbital ephemeris (T$_{\rm \pi/2}$) around the value extrapolated from \citet{Molkov2015}. For this, $T_{\rm \pi/2}$ was searched over a wide range of 0.2 d on either side of the extrapolated value in fine steps of $\sim10^{-4}$ d. We used epoch folding and $\chi^2$ maximisation technique \citep{Leahy83} for each trial ephemeris. As an example, Figure~\ref{fig:t0_Chisq} shows the variation in $\chi^2$ over the trial range of ephemeris for the \astro\ observation. This variation in $\chi^2$ was fit with a Gaussian profile over a narrow range of $\pm$0.006 d (inset of Figure~\ref{fig:t0_Chisq}). We obtained T$_{\pi/2}$ = 57630.20621(12) MJD. The T$_{\pi/2}$ determined from the other observations are listed in Table~\ref{tab:tpi}, along with the previously reported measurements. We used the bootstrap method of \citet{Boldin13} to estimate the error in the measurement of T$_{\pi/2}$. Following \citet{Sharma23}, we simulated 1000 light curves and determined the value of T$_{\pi/2}$ in each of them by the epoch folding technique. The standard deviation of T$_{\pi/2}$ measured from all the simulated light curves was taken as 1 $\sigma$ error on T$_{\pi/2}$. 

To be doubly sure and to avoid any model dependence, we varied these extrapolated values of $P_{\rm orb}$ by $\pm 2 \times 10^{-5}$ d. We did not find any significant effect of this variation on our measurement of $T_{\rm \pi/2}$.

\begin{table}
\centering
\caption{The $T_{\pi/2}$ epoch history of \lmc.}
\resizebox{\columnwidth}{!}{
\begin{tabular}{c c c c c}
\hline
Orbit $n$ & $T_{\pi/2}$ (MJD) & error (d) & Observatory &   Reference  \\
\hline
-7231 & 42829.494 & 0.019 & SAS-3 & \citep{Kelley83} \\
-4107 & 47229.3313 & 0.0004 & Ginga & \citep{Woo96} \\
-3743 & 47741.9904 & 0.0002 & Ginga & \citep{Levine91} \\
-3163 & 48558.8598 & 0.0013 & ROSAT & \citep{Woo96} \\
-2517 & 49468.6859 & 0.0054 & ASCA & \citep{Paul2002} \\
-1978 & 50227.8069 & 0.0016 & ASCA & \citep{Paul2002} \\
-1916 & 50315.12684 & 0.00008 & RXTE & Present Work \\
-1354 & 51106.6399 & 0.0025 & Beppo-SAX & \citep{Naik04} \\
-1351 & 51110.86571 & 0.00012 & RXTE$^a$ & \citep{Levine00} \\
-1351 & 51110.86600 & 0.00020 & RXTE$^b$ & \citep{Levine00} \\
-1052 & 51531.97371 & 0.00005 & RXTE & Present Work \\
-86 & 52892.46909 & 0.00045 & \xmm & Present Work \\
2200 & 56111.99880 & 0.00018 & \nus & Present Work \\
3278 & 57630.20621 & 0.00012 & \astro & Present Work \\
\hline
\multicolumn{5}{l}{$^a$ 2--8 keV.} \\
\multicolumn{5}{l}{$^b$ 8--20 keV.} \\
\end{tabular}}
\label{tab:tpi}
\end{table}

\subsection{$T_{\rm ecl}$ measurements}

\lmc\ has been continuously monitored with Burst Alert Telescope (BAT) \citep{Barthelmy05, Krimm2013} on-board the \textit{Swift} observatory \citep{Gehrels04} since 2005 and Gas Slit Camera (GSC) on-board the Monitor of All-sky X-ray Image \citep[MAXI:][]{Matsuoka2009, Mihara2011} since 2009. Therefore, in order to determine a few more mid-eclipse times, we have used the 18 year long, publicly available 15--50 keV \textit{Swift}-BAT orbital light curve\footnote{\url{https://swift.gsfc.nasa.gov/results/transients/LMCX-4/}} and 14 year long 2--20 keV \textit{MAXI}-GSC orbital light curve\footnote{\url{http://maxi.riken.jp/star_data/J0532-663/J0532-663.html}}. The photon arrival times were corrected to the solar barycenter reference time by using \textsc{earth2sun} task. 

The \emph{Swift}/BAT light curve was divided into three segments spanning about 6 years, \textit{viz.} MJD 53416--55680, 55680--57943 and 57943--60205. The \emph{MAXI}-GSC light curve was divided into two segments, \textit{viz.} MJD 55054--57032 and 57032--60212. Each of these light curves was folded at local $P_{\rm orb}$ estimated using epoch folding and the mid-eclipse time was estimated using the ramp function. Since all these time segments are relatively long ($\sim2000$ d), that the orbital period changed during these intervals. Therefore, we also folded the profiles at local $P_{\rm orb}$ and $\dot{P}_{\rm orb}$ obtained from \citet{Molkov2015}. There was no significant change in the eclipse time measurement. In fact, use of extrapolated orbital period from \citet{Molkov2015} solution resulted in a similar estimate of the mid-eclipse time, clearly indicating that the period evolution is too slow to have any considerable effect during these time intervals. 

\begin{figure}
  \centering
  \includegraphics[width=0.7\columnwidth, angle=-90]{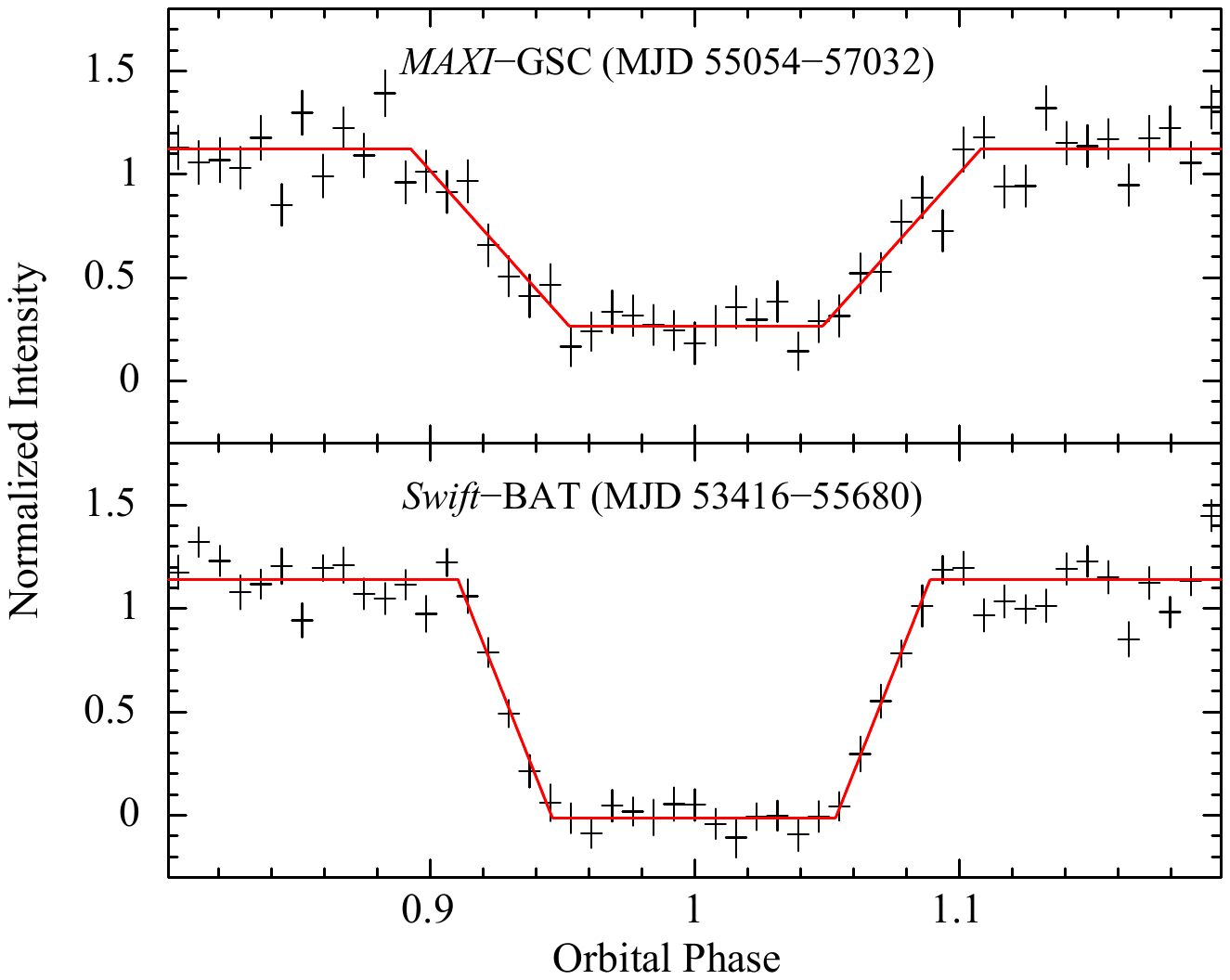}
  \caption{The eclipse profile of the light curve of \lmc\ obtained from the first segment of the long term \emph{MAXI}-GSC (top) and \textit{Swift}-BAT (below) light curves. The red solid line represents the best-fit ramp model.}
  \label{fig:batecl}
\end{figure}

Figure~\ref{fig:batecl} shows the eclipse profile of the first segment of the \emph{MAXI}-GSC and \textit{Swift}-BAT light curve fitted with a ramp function. The mid-eclipse times for these segments were found to be MJD 56042.990(3) and 54547.2978(11), respectively. Table~\ref{tab:tecl} lists the previously reported measurements of the mid-eclipse epoch ($T_{\rm ecl}$) of \lmc\ along with our estimates from \textit{Swift}-BAT and \textit{MAXI}-GSC.

\begin{table}
\centering
\caption{The mid-eclipse ($T_{\rm ecl}$) epoch history of \lmc.}
\resizebox{\columnwidth}{!}{
\begin{tabular}{c c c c l}
\hline
Orbit $n$ & $T_{\rm ecl}$ (MJD) & error (d) & Observatory &   Reference  \\
\hline
-7062 & 43067.51 & 0.02 & SAS-3 & \citep{Li78} \\
-6957 & 43215.36 & 0.02 & ESO & \citep{Chevalier1977} \\
-6772 & 43475.90 & 0.01 & CTIO & \citep{Hutchings1978} \\
-5721 & 44956.15 & 0.01 & ESO & \citep{Ilovaisky1984} \\
-5227 & 45651.917 & 0.015 & EXOSAT & \citep{Dennerl91} \\
-5224 & 45656.154 & 0.008 & EXOSAT & \citep{Pietsch1985} \\
-4662 & 46447.668 & 0.011 & EXOSAT & \citep{Dennerl91} \\
-4638 & 46481.467 & 0.003 & EXOSAT & \citep{Dennerl91} \\
-1916 & 50315.130 & 0.015 & \textit{RXTE}-PCA & \citep{Molkov2015} \\
-1614 & 50740.460 & 0.015 & \textit{RXTE}-PCA & \citep{Molkov2015} \\
-1611 & 50744.670 & 0.015 & \textit{RXTE}-PCA & \citep{Molkov2015} \\
-1349 & 51113.680 & 0.015 & \textit{RXTE}-PCA & \citep{Molkov2015} \\
-1090 & 51478.454 & 0.008 & \textit{RXTE}-ASM & \citep{Falanga15} \\
-260 & 52647.408 & 0.007 & INTEGRAL & \citep{Molkov2015} \\
-259$^a$ & 52648.804 & 0.006 & INTEGRAL & \citep{Molkov2015} \\
-86 & 52892.474 & 0.015 & XMM-Newton & \citep{Molkov2015} \\
0$^a$ & 53013.588 & 0.004 & INTEGRAL & \citep{Molkov2015} \\
2$^a$ & 53016.411 & 0.004 & INTEGRAL & \citep{Molkov2015} \\
113 & 53172.732 & 0.015 & XMM-Newton & \citep{Molkov2015} \\
887 & 54262.825 & 0.008 & \textit{RXTE}-ASM & \citep{Falanga15} \\
1089 & 54547.2978 & 0.0011 & \textit{Swift}-BAT & Present Work \\
1662 & 55354.284 & 0.009 & INTEGRAL & \citep{Molkov2015} \\
1663 & 55355.717 & 0.018 & INTEGRAL & \citep{Molkov2015} \\
1665 & 55358.531 & 0.008 & INTEGRAL & \citep{Molkov2015} \\
1678 & 55376.841 & 0.008 & INTEGRAL & \citep{Molkov2015} \\
1679 & 55378.252 & 0.008 & INTEGRAL & \citep{Molkov2015} \\
1680 & 55379.656 & 0.005 & INTEGRAL & \citep{Molkov2015} \\
1682 & 55382.463 & 0.005 & INTEGRAL & \citep{Molkov2015} \\
1684 & 55385.294 & 0.007 & INTEGRAL & \citep{Molkov2015} \\
1765 & 55499.373 & 0.006 & INTEGRAL & \citep{Molkov2015} \\
1832 & 55593.729 & 0.004 & INTEGRAL & \citep{Molkov2015} \\
1833 & 55595.130 & 0.005 & INTEGRAL & \citep{Molkov2015} \\
1834 & 55596.556 & 0.013 & INTEGRAL & \citep{Molkov2015} \\
1835 & 55597.938 & 0.006 & INTEGRAL & \citep{Molkov2015} \\
1941 & 55747.234 & 0.004 & INTEGRAL & \citep{Molkov2015} \\
1942 & 55748.645 & 0.005 & INTEGRAL & \citep{Molkov2015} \\
1944 & 55751.446 & 0.005 & INTEGRAL & \citep{Molkov2015} \\
2077 & 55938.778 & 0.009 & INTEGRAL & \citep{Molkov2015} \\
2151 & 56042.990& 0.003 & \textit{MAXI}-GSC & Present work \\
2177 & 56079.594 & 0.006 & INTEGRAL & \citep{Molkov2015} \\
2179 & 56082.424 & 0.005 & INTEGRAL & \citep{Molkov2015} \\
2198 & 56109.174 & 0.014 & INTEGRAL & \citep{Molkov2015} \\
2200 & 56111.993 & 0.006 & INTEGRAL & \citep{Molkov2015} \\
2205 & 56119.037 & 0.008 & INTEGRAL & \citep{Molkov2015} \\
2221 & 56141.583 & 0.008 & INTEGRAL & \citep{Molkov2015} \\
2222 & 56142.985 & 0.005 & INTEGRAL & \citep{Molkov2015} \\
2328 & 56292.271 & 0.006 & INTEGRAL & \citep{Molkov2015} \\
2329 & 56293.672 & 0.005 & INTEGRAL & \citep{Molkov2015} \\
2330 & 56295.084 & 0.004 & INTEGRAL & \citep{Molkov2015} \\
2438 & 56447.192 & 0.007 & INTEGRAL & \citep{Molkov2015} \\
2439 & 56448.602 & 0.006 & INTEGRAL & \citep{Molkov2015} \\
2440 & 56450.014 & 0.005 & INTEGRAL & \citep{Molkov2015} \\
2442 & 56452.824 & 0.008 & INTEGRAL & \citep{Molkov2015} \\
2460 & 56478.170 & 0.004 & INTEGRAL & \citep{Molkov2015} \\
2464 & 56483.794 & 0.008 & INTEGRAL & \citep{Molkov2015} \\
2696 & 56810.540 & 0.001 & \textit{Swift}-BAT & Present Work \\
3982 & 58621.689& 0.003& \textit{MAXI}-GSC & Present work \\
4303 & 59073.7605 & 0.0042 & \textit{Swift}-BAT & Present Work \\
\hline
\multicolumn{5}{l}{$^a$ Also reported by \cite{Falanga15}.} \\
\end{tabular}}
\label{tab:tecl}
\end{table}

\subsection{Orbital evolution}
In order to determine the orbital evolution in \lmc, we referred to a constant orbital period $P_0 = 1.40837607$ d at a reference epoch $T_0 = 53013.5878$ MJD \citep{Molkov2015}. Using these numbers, we calculated the orbital cycle ($n$) for every measurement listed in Tables~\ref{tab:tpi} and \ref{tab:tecl}. 

Initially, the orbital epochs were fit with a linear+quadratic (LQ) model to all the 58 measurements of $T_{ecl}$ and 14 measurements of $T_{\pi/2}$, using the relation, 
\begin{equation}
T_n = \Delta T_0 + \Delta P_{\rm orb} n + \frac{1}{2} \dot{P}_{\rm orb} P_{\rm orb} n^2 + \frac{1}{6} \ddot{P}_{\rm orb} P^2_{\rm orb} n^3
\label{eqn:2}
\end{equation}
The first two terms in this equation, correspond to the case of a constant orbital period. The third term describes the quadratic nature of the orbital decay in \lmc. The best fit parameters are given in Table~\ref{table:orb_evolution} (columns 4 and 5). The best-fit LQ model had reduced $\chi^2$ of 1.9 (in $T_{\rm ecl}$ measurements) and 28 (in $T_{\pi/2}$ measurements). To be more conservative and to account for much larger than 1 reduced $\chi^2$, we scaled the errors associated with $T_0$, $P_{\rm orb}$ and $\dot{P}_{\rm orb}$ by respective factor of $\sqrt{\chi^2_{\rm red}}$ \citep{Elsner80, Klis89, Iaria15}. Hereafter, this scaling of errors in $T_0$, $P_{\rm orb}$ and $\dot{P}_{\rm orb}$ has been done for all the models.  

\begin{figure}
  \centering
 \includegraphics[width=\linewidth]{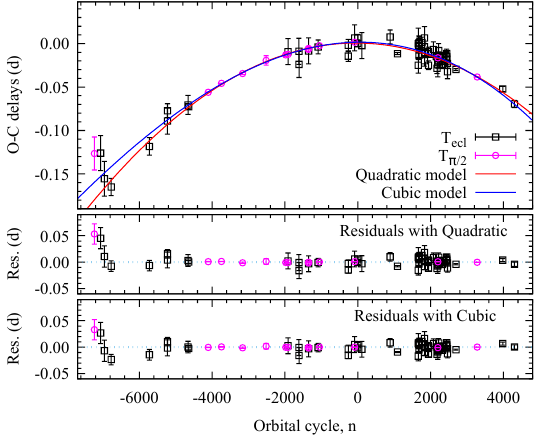}
  \caption{O-C history of \lmc\ as a function of the orbital cycle, relative to MJD 53013.5878. The data points related to $T_{ecl}$ and $T_{\pi/2}$ have been marked with black and magenta points, respectively, in the top panel. The best-fit LQ and LQC models are plotted in red and blue colours, respectively. Below two panels show the residuals with LQ and LQC models, respectively. }
  \label{fig:orbit}
\end{figure}

\begin{figure}
  \centering
 \includegraphics[width=\linewidth]{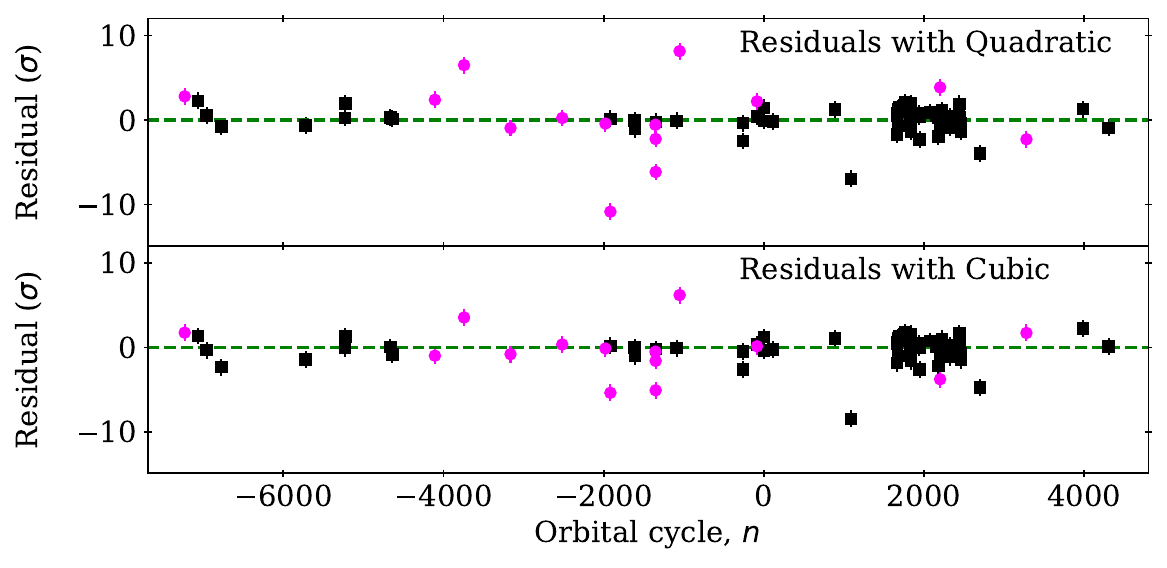}
  \caption{The residual in units of sigma (=(data-model)/error) of the O-C history of \lmc\ for LQ and LQC models, depicting the individual contribution to the net $\chi^2$. The labelling with black and magenta points is similar to that in Figure~\ref{fig:orbit}.} 
  \label{fig:orbit1}
\end{figure}

We also fitted the LQ model to the combined 72 measurements of $T_{\rm ecl}$ and $T_{\pi/2}$. The best-fit model had a $\chi^2$ of 443 for 69 degrees of freedom (d.o.f). Table~\ref{table:orb_evolution} (column 6) gives the best-fit parameters of the LQ model applied to these 72 measurements, along with their 1$\sigma$ errors (after multiplying by $\sqrt{\chi^2_{\rm red}}$, i.e. $\sim$2.5). For these combined measurements, we also fitted the linear+quadratic+cubic (LQC) model (Eqn.~\ref{eqn:2}) to the orbital epochs, where $\ddot{P}_{\rm orb}$ is the second derivative of the orbital period. The best fit had a $\chi^2$ of 296 for 68 d.o.f. The inclusion of the cubic term improved the fit statistics by $\Delta \chi^2 \sim$147 for one additional d.o.f with an $F$-test probability of $1.7\times 10^{-7}$ which corresponds to more than 3$\sigma$ significance. We found the second-period derivative to be $2.5(4)\times 10^{-13}$ d d$^{-2}$ and the evolution time scale of the period derivative ($\ddot{P}_{\rm orb}/\dot{P}_{\rm orb}$) to be $0.018 (3)$ yr$^{-1}$. Figure~\ref{fig:orbit} shows the \textit{Observed-Calculated} (O-C) residual of orbital epochs of \lmc\ relative to a constant orbital period, for both LQ and LQC models. For completeness, in order to observe the individual contribution of the quadratic and cubic components, to the net $\chi^2$, Figure~\ref{fig:orbit1} shows the residuals (in units of $\sigma$) w.r.t. LQ model (upper panel) and the LQC model (lower panel). 

\begin{table*}
\centering
\caption{Updated orbital parameters of \lmc.}
\label{table:orb_evolution}
\resizebox{2\columnwidth}{!}{
\begin{tabular}{l c c || c c c c}
\hline
       &  &    &   \multicolumn{4}{c}{Updated values: Present work$^c$}   \\
Parameter                &  \citet{Falanga15}                        &  \citet{Molkov2015}   & (using $T_{\rm ecl}$)  &  (using $T_{\pi/2}$)  &  \multicolumn{2}{c}{(using $T_{\pi/2} + T_{\rm ecl}$)} \\
& & & & & Quadratic & Cubic \\
\hline
T$_0$ (MJD)       &  53013.5910 (8) &   $53013.5878^{+0.0018}_{-0.0015}$ & $53013.5844 (14)$& $53013.5885 (3) $& $53013.58844 (14)$& $53013.5894 (2)$\\[1ex]
$P_{\rm orb}$ (d)     &  1.4083790 (7) & $1.40837607^{+4.9 \times 10^{-7}}_{-6.5 \times 10^{-7}}$ & $1.40837583 (45) $& $1.40837574 (13)$& $ 1.40837572 (6) $&  $ 1.40837655 (15) $\\[1ex]
$\dot{P}_{\rm orb}$ ($10^{-9}$ d d$^{-1}$) &  $-3.86 \pm 0.12$ &  $-4.66 \pm 0.26$ &  $ -4.60 \pm 0.19$& $ -4.97 \pm 0.08$& $-4.97 \pm  0.04$& $-5.13 \pm 0.04$\\[1ex]
$\dot{P}_{\rm orb}/P_{\rm orb}$ ($10^{-6}$ yr$^{-1}$) & -1.00 (5) & -1.21 (7) & -1.19 (5)& -1.29 (2)& -1.287 (10)& -1.33 (1)\\[1ex]
$\tau_{P_{\rm orb}}$ ($10^{6}$ yr)$^a$ & -- & -- & 0.84 & 0.77& 0.78& 0.75\\[1ex]
$\ddot{P}_{orb}$ ($10^{-13}$ d d$^{-2}$) & -- & -- & -- & -- & -- & -2.5 (4)\\[1ex]
$\ddot{P}_{orb}/\dot{P}_{orb}$ (yr$^{-1}$) & -- & -- & -- & -- & --& 0.018 (3)\\[1ex]
$\tau_{\dot{P}_{\rm orb}}$ (yr)$^b$ & -- & -- & -- & -- & --& 55\\[1ex]
\hline
\multicolumn{7}{l}{$^a$ $\tau_{P_{\rm orb}} = P_{\rm orb}/\dot{P}_{\rm orb}$ is evolution time scale of orbital period.}\\
\multicolumn{7}{l}{$^b$ $\tau_{\dot{P}_{\rm orb}} = \dot{P}_{\rm orb}/\ddot{P}_{\rm orb}$ is evolution time scale of orbital period derivative.}\\
\multicolumn{7}{l}{$^c$ Error in T$_0$, P$_{\rm orb}$ and $\dot{P}_{\rm orb}$ have been artificially increased such that respective $\chi^2_{red}\sim$1}

\end{tabular}}
\end{table*}

\begin{figure*}
  \centering
 \includegraphics[width=0.9\linewidth]{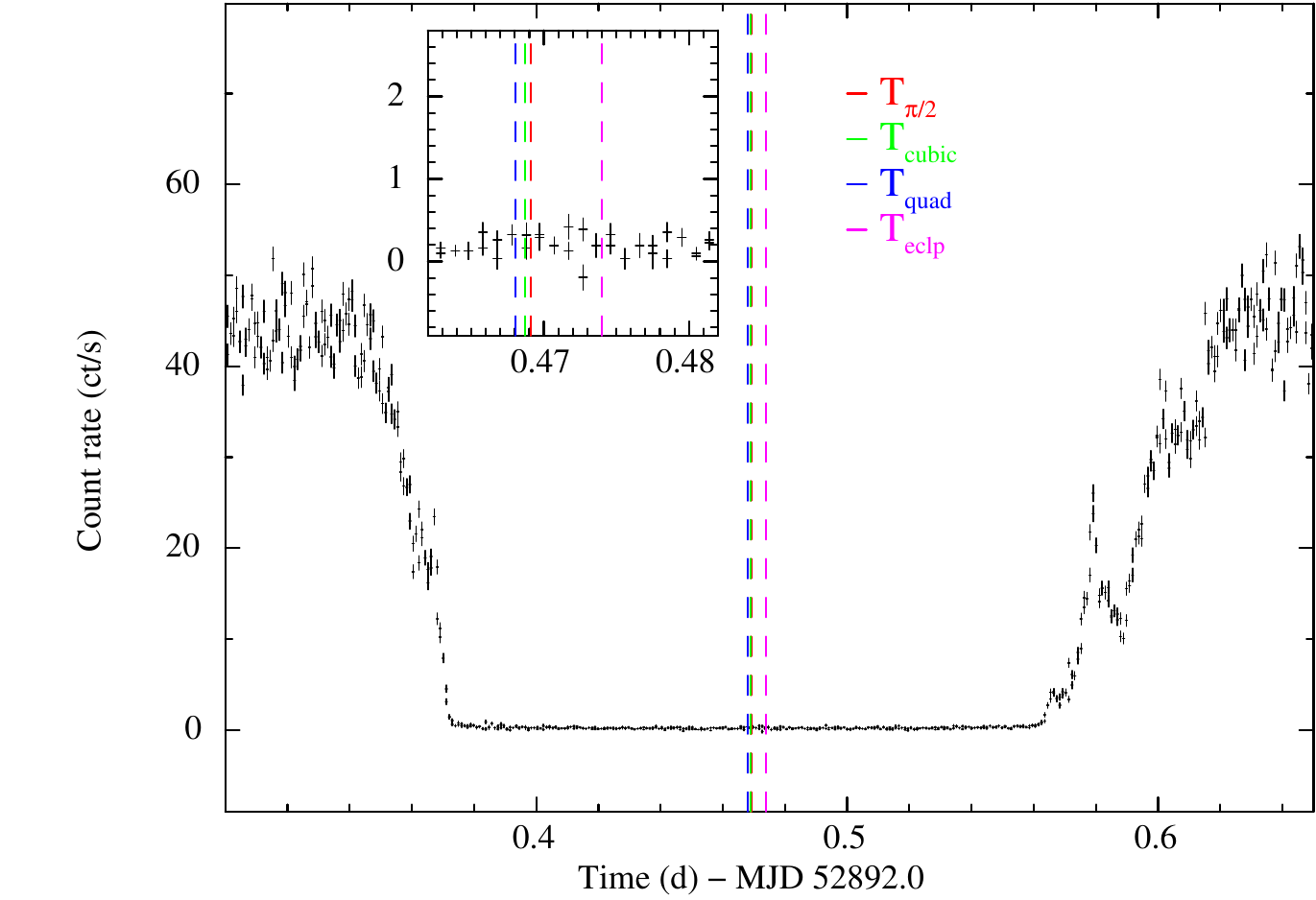}
  \caption{Eclipse profile of the \xmm\ observation (Obs. Id. 0142800101) used in the current work. The vertical lines correspond to the expected mid-eclipse time from various techniques. Red: $T_{\pi/2}$, Blue: T$_n$ from LQ model, Green: T$_n$ from LQC, and Magenta: T$_{ecl}$ from \citet{Molkov2015}. The inset figure shows the zoomed-in section near the centre of the eclipse.}
  \label{fig:eclxmm}
\end{figure*}

Amongst the observations with the narrow field instruments analysed in this work, only the observation from \xmm\ covered a complete X-ray eclipse of \lmc. The results from the eclipse timing are already reported in \citet{Molkov2015}. However, the errors quoted in \citet{Molkov2015} are as large as 1300 s. This is perhaps due to the post-eclipse dips, which can affect the determination of the eclipse egress profile. The difference between the two measurement techniques and the two orbital evolution models, with and without a second derivative of the orbital period is shown in Figure \ref{fig:eclxmm}. The vertical dashed lines in this figure correspond to mid-eclipse time measurement obtained from the eclipse timing technique (magenta color) by \citet{Molkov2015}, $T_{\pi/2}$ measured from the pulse timing (reported in Table~\ref{tab:tpi}) in red color, $T_n$ calculated from the best-fit LQ model (blue color) and LQC model (green color). The $T_{\pi/2}$ measurement is closest to the estimates from the LQC model. This inference also supports the fact the pulse arrival time analysis is much more accurate than the mid-eclipse measurements.

\section{DISCUSSION}

In this paper, we have revisited a very well-studied HMXB system, \lmc\ and have made robust estimates of its orbital evolution, by adding new measurements from light curves obtained with the pointed observations made with \RXTE, \xmm, \nus, \astro, and long term light curves obtained with \textit{Swift}-BAT and \textit{MAXI}-GSC data and thereby spanning the ephemerides history over almost 45 years. We have determined an orbital period of 1.40837655(15) d at MJD 53013.5894(2), decaying at a rate of -5.13(4)$\times$10$^{-9}$ d d$^{-1}$, with an orbital evolution timescale of about 0.8 Myr. We have also estimated a second derivative of orbital period $2.5(4)\times 10^{-13}$ d d$^{-2}$, implying that the orbital decay rate is evolving at a time scale of just about 55 years. We discuss these results in the light of,
\begin{enumerate}
    \item Orbital evolution mechanisms at play in \lmc.
    \item Importance of existence of $\ddot{P}_{orb}$.
    \item Importance of a non-zero $e \cos{\omega}$ term.
\end{enumerate}

Orbital evolution has been found to occur in a handful of low-mass X-ray binaries (LMXBs). Amongst accretion-powered millisecond X-ray pulsars (AMXPs), a rapid orbit expansion has been observed in SWIFT J1749.4--2807 \citep{Sanna22},  SAX J1808.4--3658 \citep{Jain07, Burderi09, Illiano2023} and IGR J17062--6143 \citep{Bult21}. In few other AMXPs, only the upper limits on the orbital period derivatives are known \citep{Bult18, Sanna18, Sanna22}, and all are compatible with a rapidly expanding orbit. Amongst eclipsing LMXBs and/or slowly rotating pulsars, orbital evolution timescale is known for X2127+119 \citep{Homer98}, 4U 1822--37 \citep{Parmar00, Jain10b}, AX J1745.6--2901 \citep{Ponti2017}, MXB 1658--298 \citep{Jain17}, Her X--1 \citep{Deeter91, Staubert09}, 4U 1820--303 \citep{Klis93, Peuten14}, EXO 0748--676 \citep{Wolff09} and XTE J1710--281 \citep{Jain11, Jain22}. The orbital evolution in these systems is known to be positive (in X2127+119 and 4U 1822--37), negative (in AX J1745.6--2901, MXB 1658--298, Her X--1 and 4U 1820--303), and has shown sudden changes (again, both positive and negative) in period (for example, EXO 0748--676 and XTE J1710--281), likely due to magnetic cycling in the companion star. The increasing orbital separation in these systems has been attributed to short-lived mass exchange episodes and strong tidal interaction between the binary components. MXB 1658--298 shows an overall orbital period decay influenced by the presence of a massive circumbinary planet. The cause of orbital decay in most LMXBs is inconclusive because the observed time scale of evolution in these systems (about a million years) is too fast for a conservative Roche lobe mass transfer.

In case of HMXBs, orbital evolution has been found in several systems like, Cen X--3 \citep{Kelley83b, Raichur10, Klawin23}, \lmc\ \citep{Levine00, Naik04, Molkov2015}, SMC X--1 \citep{Levine93, Wojdowski98, Raichur10}, OAO 1657--415 \citep{Jenke12}, 4U 1700--37 \citep{Rubin96, Falanga15, Islam2016}, 4U 1538--52 \citep{Baykal06, Mukherjee06}, GX 301--2 \citep{Doroshenko10} and Cyg X--3 \citep{Singh02, Bhargava17}. Orbits of all these binary systems are known to decay, except Cyg X--3. The orbital evolution in Cen X--3, \lmc, SMC X--1 and OAO 1657--415 is predominantly due to tidal interactions between the binary components and/or transfer of angular momentum due to strong stellar wind from the binary system to a halo surrounding it. The rate of orbital period decay is much smaller in the case of 4U 1700--37 and 4U 1538--52 than the remaining HMXBs. 

Based on the orbital modulation, several authors have studied the changes in the orbital period of Cyg X--3 \citep{Klis89, Kitamoto92, Kitamoto95, Singh02, Bhargava17}. Some earlier works based on data before 1990 \citep{Klis89, Kitamoto92}, reported a second derivative of the orbital period but detection was marginal and results were inconclusive. The detection significance of $\ddot{P}_{\rm orb}$ crucially depended on the intrinsic scatter in the data. The inclusion of more data (up to 1993) gave a smaller value $\ddot{P}_{\rm orb}$ \citep{Kitamoto95}. It was even smaller from data up to 2001 \citep{Singh02}. The cubic fit to the O-C curve was only marginally better than quadratic fit. The latest works by \citet{Bhargava17} over 45 years of time base are biased towards a secular variation in the orbital period without any requirement of a second derivative. They have however hinted towards short-term local period variations linked with jet emission. 

Assuming a conservative mass transfer in \lmc, where the neutron star accretes all the matter lost by its companion, the rate of change of the orbital period is given by \citep{Heuvel73},
\begin{equation}
    \dot{P}_{\rm orb}/P_{\rm orb} = 3 \frac{(M_c-M_{NS})}{M_c M_{NS}} \dot{M}_c
    \label{eqn:discussion1}
\end{equation}
This gives $\dot{M}_c \sim -7.6 \times 10^{-7} M_\odot$yr$^{-1}$, assuming orbital decay rate of $\dot{P}_{\rm orb}/P_{\rm orb} = -1.31 \times 10^{-6}$  yr$^{-1}$,  neutron star mass of $M_{NS} = 1.57 M_\odot$, and a companion mass of $M_{NS} = 18 M_\odot$ \citep{Falanga15}. This estimate is nearly 3 times more than the theoretical mass loss estimate of $\sim -2.4 \times 10^{-7} M_\odot$yr$^{-1}$ \citep{Falanga15}; and it exceeds the Eddington mass accretion limit by a large factor. Clearly, other mass loss mechanisms are at work in addition to conservative mass transfer and tidal decay.

Assuming a non-conservative mass transfer, where only a fraction of the ejected matter is accreted by the neutron star; following \citet{Heuvel94} and \citet{Jenke12}, and referring to orbital parameters of \lmc, we have determined a lower limit on the angular momentum transferred through stellar winds from the companion by using, 
\begin{equation}
    - \dot{P}_{\rm orb}/P_{\rm orb} = - (1+3\gamma) \frac{\dot{M}_c}{M_c + M_{NS}} -3 \frac{\dot{M}_c}{M_c}
\end{equation}
where $\gamma$ is the ratio of escaping angular momentum per unit mass to the total angular momentum per unit mass and given by 
\begin{equation}
    \gamma = \frac{(M_c + M_{NS})^2}{M_c ~M_{NS}} \Big( \frac{a_e}{a(1-e^2)} \Big)^{1/2}
\end{equation}
where $a$ is the semi-major axis of the orbit and $a_e$ is the radius beyond the L$_2$ point of the system of escaping material. Assuming $a_e > 1.2 a$, and neutron star mass of $M_{NS} = 1.57 M_\odot$, a companion mass of $M_{NS} = 18 M_\odot$ and mass loss rate of $M_c\sim -2.4 \times 10^{-7} M_\odot$yr$^{-1}$ \citep{Falanga15}, we get a decay rate of  $>$ --1.4$\times$10$^{-6}$ yr$^{-1}$. This is well consistent with the observed $\dot{P}_{orb}/P_{orb}$ value of --1.3$\times$10$^{-6}$ yr$^{-1}$. 

For most binary systems, one can measure the period derivative and a first order estimate for the orbital evolution timescale. However, if the period derivative itself changes over time, then this estimation is inaccurate. We, for the first time, have determined a second derivative of the orbital period, which turns out to be quite short. So the period evolution timescale that is measured today is perhaps not valid in a few decades. If the same is true for the other HMXBs, then the long-term evolution of HMXBs can not be estimated from the current measurements of $P_{\rm orb}$ and $\dot{P}_{\rm orb}$. However, \lmc\ should be observed extensively over the next decade to have a more secure determination of the second period derivative.

For eccentric orbits, $T_{\rm ecl}$ (measured from the eclipse timing technique) generally does not coincide with $T_{\pi/2}$ (determined from the pulse arrival time technique). This time delay, to the first order in eccentricity, is given by 
\begin{equation}
T_{\pi/2} = T_{\rm ecl} + \frac{eP_{orb}}{\pi} \cos{\omega}
\label{eqn:ecc}
\end{equation}

where $\omega$ is the argument of periapsis \citep{Falanga15}. In case of \lmc, the difference in the best-fit value of T$_0$ from T$_{ecl}$ and T$_{\pi/2}$ measurements is 0.0041 (14) d, which implies a value of $\sim$0.009 (3) for $e\cos{\omega}$. 

The tidal interaction between the binary components of HMXBs is known to circularize their orbit. As a result, the systems amongst these with small orbital periods (of up to 4 days) have very low eccentricity. Updated orbital solution for HMXBs is important because a good estimate of eccentricity of the orbit has a potential to constrain the age of the binary system and also serve as indirect detection of gravitational wave emission from the system. Likewise, a good estimate of the periastron angle and advancement of the periastron angle with time \citep[e.g., 4U 0115+63:][]{Raichur2010b} is a clue to understanding the stellar structure models of these massive binary systems. Our estimate of $e\cos{\omega}$ in the case of \lmc\ can serve as an important ingredient in determining these numbers.

\section*{Acknowledgements}

We thank the anonymous referee for insightful comments and suggestions. This research has made use of archival data and software provided by NASA’s High Energy Astrophysics Science Archive Research Center (HEASARC), which is a service of the Astrophysics Science Division at NASA/GSFC. This work has made use of data from the \astro\ mission of the Indian Space Research Organisation (ISRO), archived at the Indian Space Science Data Centre (ISSDC). We  thank the LAXPC Payload Operation Center (POC) at TIFR, Mumbai for providing necessary software tools. This research has also used data from \textit{Swift}-BAT and \textit{MAXI}-GSC. 

\section*{Data Availability}

Data used in this work can be accessed through the Indian Space Science Data Center (ISSDC) at 
\url{https://astrobrowse.issdc.gov.in/astro\_archive/archive/Home.jsp} and HEASARC archive at \url{https://heasarc.gsfc.nasa.gov/cgi-bin/W3Browse/w3browse.pl}.



\bibliographystyle{mnras}
\bibliography{ref} 







\bsp	
\label{lastpage}
\end{document}